\documentclass[
  aps,
  prd,
  twocolumn,
  superscriptaddress,
  nofootinbib,
  amsmath,amssymb,
  floatfix
]{revtex4-2}

\usepackage{graphicx}
\usepackage{dcolumn}
\usepackage{bm}
\usepackage{hyperref}
\usepackage{color}

\begin{document}

\title{Antikick Relation in High-Energy Head-On
Collisions of Spinning Black Holes}

\author{Carlos O. Lousto}
\affiliation{Center for Computational Relativity and Gravitation,
Rochester Institute of Technology, Rochester, New York 14623, USA}
\author{James Healy}
\affiliation{Center for Computational Relativity and Gravitation,
Rochester Institute of Technology, Rochester, New York 14623, USA}
\author{Alessandro Ciarfella}
\affiliation{Center for Computational Relativity and Gravitation,
Rochester Institute of Technology, Rochester, New York 14623, USA}
\author{Hiroyuki Nakano}
\affiliation{Faculty of Law, Ryukoku University, Kyoto 612-8577, Japan}

\date{\today}

\begin{abstract}
The collision of black holes at relativistic speeds probes gravity in its
most extreme dynamical regime. While the maximum gravitational recoil from
\emph{grazing} high-energy collisions ($\approx28\,562$~km/s, i.e., $\sim0.1c$)
and the maximum radiated energy $E_{\rm rad}$ and remnant spin $\alpha_f^{\max}$ from such encounters
($E_{\rm rad}/M_{\rm ADM}\approx32\%$ where $M_{\rm ADM}$ is the ADM mass, 
and $\alpha_f^{\max}\approx0.987$) have
been established previously~\cite{Healy:2022jbh,Healy:2024lhl}, here we focus
on the \emph{head-on} high-energy collision of equal-mass spinning black
holes and on the detailed structure of the resulting recoil. Performing a
sequence of full numerical simulations for spin magnitudes $s=0.5,0.65$,
and $0.8$ over a range of initial momenta $\gamma v$, we characterize the peak
recoil $V_p$, the final recoil $V_f$, and the antikick $\Delta V\equiv V_f-V_p$, and we
provide phenomenological fits of their dependence on $\gamma v$ and $s$. We
complement these results with a zero-frequency-limit (ZFL) analysis of the
radiated energy and momentum, a quasinormal-mode model of the antikick, and a superposed boosted
double-Kerr close-limit estimate. We find that in the relativistic regime ($\gamma v>1$) the peak and final
recoil are directly proportional, $V_p\approx7.4\,V_f$ (equivalently
$\Delta V \approx-6.4\,V_f$), largely independent of both the
initial momentum and the spin magnitude, pointing to a common post-merger
relaxation. While the ZFL predicts a leading linear-in-spin dependence, the close-limit analysis predicts a leading $s^3$
dependence of the recoil amplitude; with the three spin magnitudes studied
here the empirical exponent is $s^{1.27\pm0.08}$, motivating an even
higher energy collision spin sequence study.
\end{abstract}

\maketitle

\section{Introduction}
The high-energy collision of two black holes probes general relativity in
its most extreme dynamical regime, with applications to the gauge/gravity
duality and holography~\cite{Berti:2010ce}, to tests of the radiation-bound
theorems and the cosmic censorship conjecture~\cite{Smarr:1977fy}, and to
the dynamics of primordial black holes in the early Universe. The maximum
energy radiated in the head-on collision of two equal-mass, nonspinning
black holes was first estimated with full numerical relativity by
Sperhake \textit{et al.}~\cite{Sperhake:2008ga}, who found an efficiency of
$14\pm3\%$, and was subsequently argued to be largely independent of the
black-hole spin~\cite{Sperhake:2012me}. Using new initial data with reduced
spurious radiation content~\cite{Ruchlin:2014zva}, which allow boosts up to
$v\sim0.99c$, Ref.~\cite{Healy:2015mla} revisited the nonspinning head-on
problem and placed the maximum radiated energy at $\approx13\%\pm1\%$. Grazing
high-energy collisions were studied in Ref.~\cite{Sperhake:2009jz}, giving
$\approx35\%$ at a critical impact parameter, and in notable analytic detail
within the zero-frequency-limit (ZFL) framework in Ref.~\cite{Berti:2010ce}.

More recently, large numerical campaigns were carried out mapping the extremes of grazing encounters. Reference
\cite{Healy:2022jbh} performed $1381$ simulations of equal-mass binaries
with opposite in-plane spins, searching the four-dimensional parameter space
of momentum $\gamma v$, impact parameter $b$, spin orientation $\varphi$, and
spin magnitude $s$, and extrapolating to the extremal limit $s\to1$ a maximum
recoil of $28\,562\pm342$~km/s, i.e., approximately $10\%$ of the speed of
light. Reference~\cite{Healy:2024lhl} then performed $769$ simulations with
spins aligned with the orbital angular momentum, finding a maximum radiated
energy $E_{\rm rad}/M_{\rm ADM}\approx32\%\pm2\%$ where $M_{\rm ADM}$ is the ADM mass
(weakly dependent on the intrinsic spin) 
and a remnant spin as large as $\alpha_f^{\max}=0.987$,
consistent with cosmic censorship.

Whereas those works targeted the global \emph{maxima} of the recoil and of
the radiated energy in grazing configurations, here we turn to the
complementary problem of the \emph{head-on} high-energy collision of spinning
black holes and the detailed \emph{structure} of the recoil it produces. In
a head-on encounter the net linear momentum radiated vanishes for nonspinning
equal-mass holes, so any recoil is intrinsically a spin effect; moreover, the
recoil is generated almost entirely after merger, during the ringdown, and
exhibits a pronounced antikick in which the velocity peaks at $V_p$ and then
relaxes to a smaller final value $V_f$. We quantify $V_p$, $V_f$, and the
antikick $\Delta V \equiv V_f-V_p$ as functions of $\gamma v$ and $s$, and we interpret them
analytically through (i) a ZFL expansion of the radiated
energy and momentum (Appendix~\ref{app:zfl}), (ii) a quasinormal-mode (QNM) model
of the post-merger ringdown (Appendix~\ref{app:qnm}), and (iii) a superposed
boosted double-Kerr close-limit estimate (Appendix~\ref{app:cl}).
The close-limit analysis predicts a leading $s^3$ dependence of the recoil;
the three spin magnitudes studied here in the $\gamma v\leq3.2$ 
give $s^{1.3\pm0.1}$ for the
empirical exponent, and reveal a robust, nearly universal
relation between the peak and final recoil.

\section{Numerical methods and parameter space}
We evolve the binaries with the \textsc{LazEv}~\cite{Zlochower:2005bj}
implementation of the moving-puncture approach~\cite{Campanelli:2005dd}
using the BSSNOK formalism~\cite{Shibata95,Baumgarte99}, with eighth-order
spatial finite differencing, a fourth-order Runge--Kutta time integrator at Courant factor
$\Delta t/\Delta x=1/3$,
 and the \textsc{Cactus}/\textsc{Carpet}
moving-boxes mesh refinement. Initial data are Bowen--York puncture data
generated with the \textsc{TwoPunctures} spectral solver~\cite{Ansorg:2004ds};
for the highest boosts, the reduced-spurious-radiation data of
Ref.~\cite{Ruchlin:2014zva} can be used to reach $v\to c$. Apparent horizons
are located with \textsc{AHFinderDirect}~\cite{Thornburg2000:multiple-patch-evolution},
and the horizon mass $m_H$ and spin $S_H$ are measured with the isolated-horizon
algorithm~\cite{Dreyer:2002mx,Campanelli:2006fy}, with $m_H^2=m_{\rm irr}^2+S_H^2/(4m_{\rm irr}^2)$.
The radiated energy, linear momentum, and angular momentum are computed from
the Weyl scalar $\psi_4$~\cite{Campanelli99,Lousto:2007mh} extracted at a finite
observer radius and extrapolated to null infinity with the perturbative
formulas of Ref.~\cite{Nakano:2015pta}.

\begin{figure}[t]
  \includegraphics[width=0.8\columnwidth]{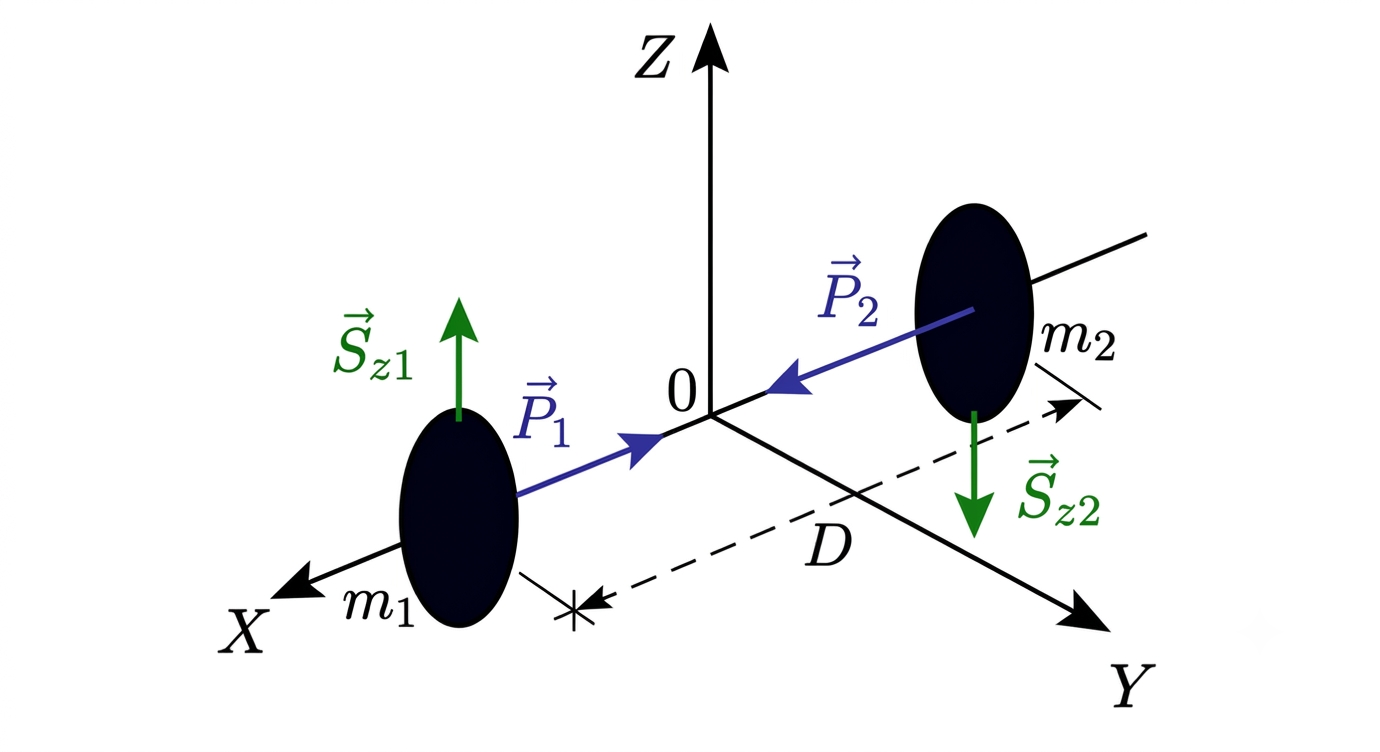}
  \caption{\label{fig:config}
Schematic representation of the initial state for the high-energy, head-on collision of two equal-mass, opposite-spin and opposite-momentum black holes. The system is depicted in a 3D cartesian coordinate system (X, Y, Z) centered at the origin (0). The two black holes, $m_1$ and $m_2$, are represented by dark spheroids. Green vectors $\vec{S}_{z1}$ and $\vec{S}_{z2}$ indicate that the black holes possess opposite spins, which are anti-aligned along the Z-axis. Blue vectors $\vec{P}_1$ and $\vec{P}_2$ represent their opposite linear momenta, which are directed head-on along the collision axis toward the origin. The configuration depicts the moment of initial release. Both black holes have equal mass, $m_1 = m_2$. The initial separation distance, indicated by the dashed dimension line labeled $D$, is up to $D = 400M$ (increasing with $\gamma v$).}
\end{figure}

The configurations are equal-mass binaries, $m_1=m_2$, colliding head-on
along the $x$ axis with anti-aligned spins $\vec s_1=-\vec s_2$ of magnitude
$s=|\vec S_H|/m_H^2$ oriented along the $z$ axis, perpendicular to the line of
collision, as sketched in Fig.~\ref{fig:config}. This opposite-spin arrangement breaks the reflection symmetry across the
collision plane ($y\to-y$) that would otherwise forbid a net recoil for
equal masses, generating a kick along the $y$ axis, perpendicular to both
the spins and the collision direction, following the vector product
$\vec v\times\vec S$ of the leading spin-orbit
flux~\cite{Racine:2008kj}. (In the perturbative analyses of the Appendices,
the axes are rotated so that the recoil lies along the polar
axis, and its component is there generically denoted $P_z$.)
We label
the incoming boost by the momentum per horizon mass $\gamma v\equiv p/m_H$,
with $\gamma=(1-v^2)^{-1/2}$, and scan $\gamma v$ from $0.3$ to $3.2$ for each
spin magnitude $s=0.5,\,0.65$, and $0.8$. The holes are started from large initial
separations (up to $D=400M$, increasing with $\gamma v$) so that the boost
parameter corresponds closely to the asymptotic momentum and the spurious
initial-data radiation can be cleanly separated. The recoil velocity is extracted from the time integral of the radiated
linear-momentum flux $dP_y/dt$, after subtracting the initial spurious Bowen--York radiation burst in
the time domain; we record both the signed recoil velocity at peak magnitude,
$V_p\equiv V(t_p)$ with $t_p=\mathrm{arg\,max}_t|V(t)|$, and the signed,
relaxed final value $V_f$, and define the antikick as
$\Delta V\equiv V_f-V_p$ (see the inset of Fig.~\ref{fig:dv_gv}); in the
relativistic regime $V_p$ and $V_f$ carry the same sign and
$|V_p|>|V_f|$, so $\Delta V>0$.

Following the convergence studies of
Refs.~\cite{Healy:2022jbh,Healy:2024lhl}, where finite-difference errors
were assessed with resolutions increased in steps of $1.2$ ($n100$, $n120$,
and $n144$), we adopt $n100$ grids for $\gamma v<2.0$ and $n150$ grids for the
higher-momentum runs. Both use 12 levels of mesh refinement centered on
each puncture, supplemented by a set of fixed levels centered on the
collision point that cover the merger region. The $n100$ grids place the
outer boundaries at $800M$ with a coarse-grid spacing of $8M$ and a finest
spacing of $M/256$, while the $n150$ grids place the boundaries at $600M$
with a coarse spacing of $4M$ and a finest spacing of $M/512$, i.e., $50\%$
more points across the coarse and intermediate levels and twice the
resolution on the levels covering the horizons. The Weyl scalar $\psi_4$
is decomposed into spin-weighted spherical harmonics up to $\ell=6$ on ten
extraction spheres between $r=75M$ and $r=375M$, and the radiative
quantities (radiated energy and linear momentum) are extrapolated to null
infinity with the perturbative formulas of Ref.~\cite{Nakano:2015pta},
after subtracting the initial burst of spurious radiation in the time
domain.

As an internal consistency check, the measured initial irreducible masses
$m_H^i$ listed in
Tables~\ref{tab:kinematics_spin50}--\ref{tab:kinematics_spin80} agree with
the analytic boosted-puncture value $1/(2\gamma)$,
$\gamma=\sqrt{1+(\gamma v)^2}$, at the percent level (sub-percent through
$\gamma v\lesssim2$), the small residual being the initial binding energy
and the finite resolution of highly boosted horizons, and are consistent
with the published nonspinning head-on sequence~\cite{Healy:2015mla};
likewise, the radiated energies in
Tables~\ref{tab:kinematics_spin50}--\ref{tab:kinematics_spin80}, which
depend only weakly on $s$, agree with that nonspinning sequence at the
corresponding $\gamma v$.

\section{Results}

\subsection{Tables of results}
%


\begin{table}[!htb]
\caption{\label{tab:kinematics_spin50} Radiated energy and recoil velocities
for head-on collisions with dimensionless spins $s_{1,2} = \pm0.50$. The
table details the initial boost parameter $\gamma v$, the measured initial irreducible mass of each hole, $m_H^i$ (analytically
$1/(2\gamma)$ with $\gamma=\sqrt{1+(\gamma v)^2}$ and the total rest mass
normalized to unity), the radiated gravitational wave energy
$E_{\rm rad}/M$ (normalized by the initial mass), the final remnant recoil
velocity $V_f$, and the signed peak recoil velocity $V_p$ reached during
the ringdown.
All radiative quantities are extrapolated to extraction at infinity.}
\begin{ruledtabular}
\begin{tabular}{ccccc}
$\gamma v$ & $m_H^i$ & $E_{\rm rad}/M$ & $V_f$ (km/s) & $V_p$ (km/s) \\
\colrule
0.300 & 0.4790 & 0.001118 & $-11.56$ & $7.20$ \\
0.500 & 0.4474 & 0.003073 & $-6.30$ & $14.61$ \\
0.800 & 0.3909 & 0.010465 & $0.82$ & $25.76$ \\
1.000 & 0.3542 & 0.017398 & $4.36$ & $27.25$ \\
1.200 & 0.3209 & 0.024896 & $1.60$ & $-2.53$ \\
1.500 & 0.2784 & 0.035172 & $-1.37$ & $-39.55$ \\
1.800 & 0.2440 & 0.043855 & $-7.15$ & $-94.91$ \\
2.000 & 0.2249 & 0.052173 & $-14.45$ & $-148.76$ \\
2.400 & 0.1937 & 0.060240 & $-29.54$ & $-239.48$ \\
2.800 & 0.1695 & 0.065566 & $-43.27$ & $-321.13$ \\
3.000 & 0.1595 & 0.067206 & $-53.25$ & $-374.73$ \\
3.200 & 0.1505 & 0.069095 & $-53.77$ & $-392.24$ \\
\end{tabular}
\end{ruledtabular}
\end{table}

\begin{table}[!htb]
\caption{\label{tab:kinematics_spin65} Radiated energy and recoil velocities for head-on collisions with dimensionless spins of $s_{1,2} = \pm0.65$. The table details the initial boost parameter $\gamma v$, the measured initial irreducible mass $m_H^i$ (as in
Table~\ref{tab:kinematics_spin50}), the radiated gravitational wave energy
$E_{\rm rad}/M$ (normalized by the initial mass), the final remnant recoil
velocity $V_f$, and the signed peak recoil velocity $V_p$ reached during
the ringdown. All radiative quantities are extrapolated to extraction at infinity.}
\begin{ruledtabular}
\begin{tabular}{ccccc}
$\gamma v$ & $m_H^i$ & $E_{\rm rad}/M$ & $V_f$ (km/s) & $V_p$ (km/s) \\
\colrule
0.300 & 0.4789 & 0.001124 & $-15.18$ & $8.49$ \\
0.500 & 0.4474 & 0.003079 & $-9.03$ & $12.45$ \\
0.800 & 0.3911 & 0.010459 & $0.83$ & $29.32$ \\
1.000 & 0.3544 & 0.017365 & $3.35$ & $19.92$ \\
1.200 & 0.3212 & 0.024757 & $0.47$ & $-15.78$ \\
1.500 & 0.2787 & 0.035296 & $-11.99$ & $-107.74$ \\
1.800 & 0.2428\footnotemark[1] & 0.043552 & $-10.35$ & $-136.84$ \\
2.000 & 0.2253 & 0.051943 & $-23.51$ & $-219.53$ \\
2.400 & 0.1923\footnotemark[1] & 0.059935 & $-42.63$ & $-334.71$ \\
2.800 & 0.1682\footnotemark[1] & 0.065220 & $-57.10$ & $-425.75$ \\
3.000 & 0.1599 & 0.067222 & $-65.80$ & $-481.67$ \\
3.200 & 0.1491\footnotemark[1] & 0.068609 & $-72.00$ & $-521.97$ \\
\end{tabular}
\end{ruledtabular}
\footnotetext[1]{Horizon-finder measurement unavailable for this run;
the analytic estimate $1/(2\gamma)$ is quoted.}
\end{table}

\begin{table}[!htb]
\caption{\label{tab:kinematics_spin80} Radiated energy and recoil velocities for head-on collisions with dimensionless spins of $s_{1,2} = \pm0.80$. The table details the initial boost parameter $\gamma v$, the measured initial irreducible mass $m_H^i$ (as in
Table~\ref{tab:kinematics_spin50}), the radiated gravitational wave energy
$E_{\rm rad}/M$ (normalized by the initial mass), the final remnant recoil
velocity $V_f$, and the signed peak recoil velocity $V_p$ reached during
the ringdown. All radiative quantities are extrapolated to extraction at infinity.}
\begin{ruledtabular}
\begin{tabular}{ccccc}
$\gamma v$ & $m_H^i$ & $E_{\rm rad}/M$ & $V_f$ (km/s) & $V_p$ (km/s) \\
\colrule
0.300 & 0.4788 & 0.001138 & $-19.72$ & $2.16$ \\
0.500 & 0.4475 & 0.003131 & $-11.16$ & $16.90$ \\
0.800 & 0.3913 & 0.010268 & $-1.26$ & $14.78$ \\
1.000 & 0.3548 & 0.017270 & $-0.92$ & $-9.54$ \\
1.200 & 0.3217 & 0.024567 & $-4.10$ & $-51.98$ \\
1.500 & 0.2793 & 0.035180 & $-4.66$ & $-98.04$ \\
1.800 & 0.2450 & 0.043413 & $-31.24$ & $-253.62$ \\
2.000 & 0.2259 & 0.051523 & $-38.09$ & $-320.66$ \\
2.400 & 0.1946 & 0.059456 & $-58.40$ & $-450.17$ \\
2.800 & 0.1705 & 0.064635 & $-77.38$ & $-572.36$ \\
3.000 & 0.1581\footnotemark[1] & 0.066614 & $-89.33$ & $-631.79$ \\
3.200 & 0.1491\footnotemark[1] & 0.068239 & $-101.92$ & $-704.45$ \\
\end{tabular}
\end{ruledtabular}
\footnotetext[1]{Horizon-finder measurement unavailable for this run;
the analytic estimate $1/(2\gamma)$ is quoted.}
\end{table}

\subsection{Recoil structure: peak, final, and antikick}
\begin{figure}[t]
  \includegraphics[width=\columnwidth]{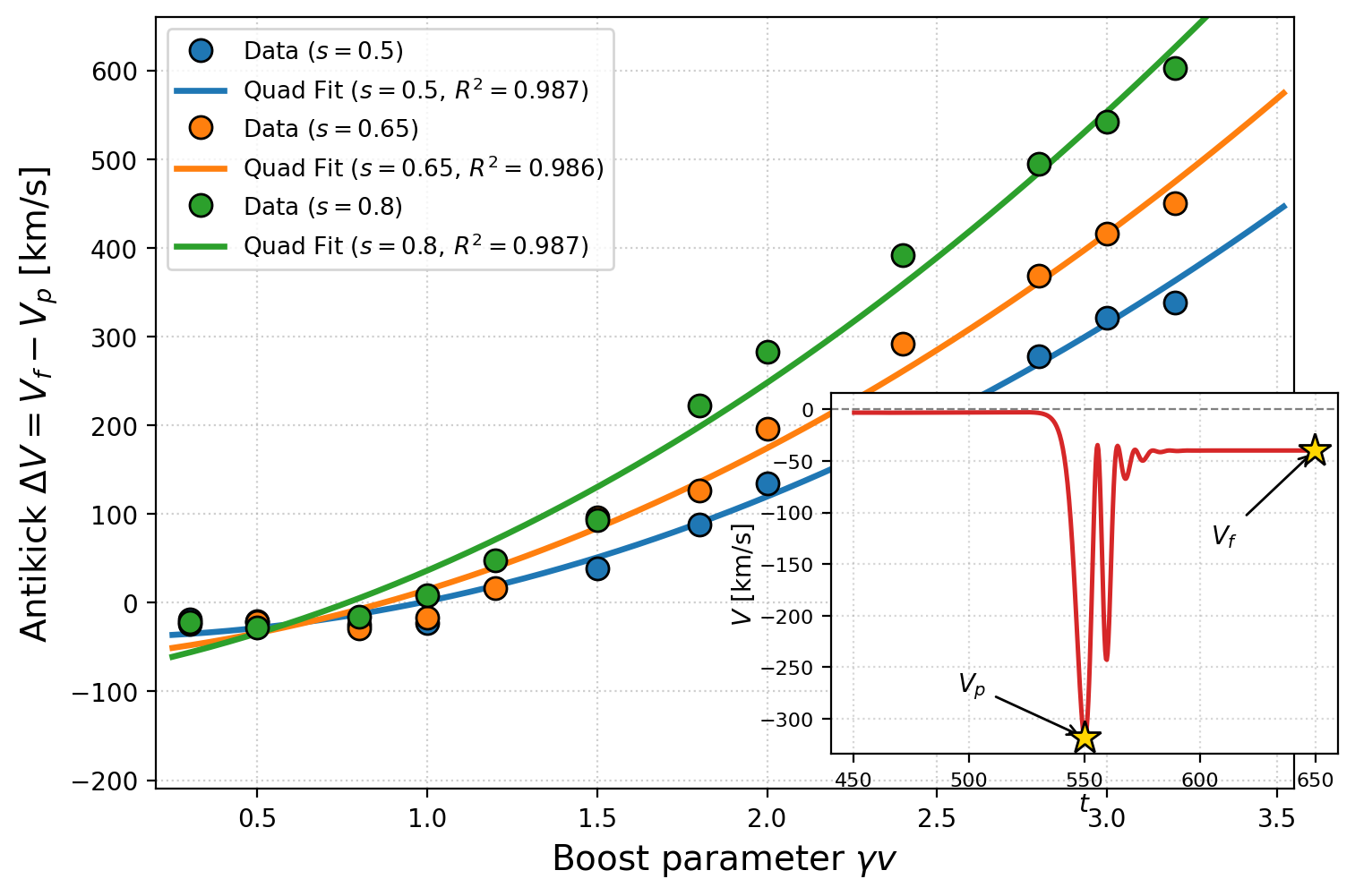}
  \caption{\label{fig:dv_gv}
Main panel: The antikick velocity $\Delta V \equiv V_f - V_p$ (in km/s) generated during the ringdown phase of the remnant black hole as a function of the initial boost parameter $\gamma v$. Results are shown for three distinct dimensionless spin magnitudes of the primary black hole: $s = 0.50$ (blue), $s = 0.65$ (orange), and $s = 0.80$ (green). The solid curves represent quadratic fits in $\gamma v$, demonstrating excellent agreement ($R^2\approx0.99$) with the strongly relativistic data. Inset: The time evolution of the cumulative recoil velocity $V(t)$ (in
km/s, from the integrated $dP_y/dt$ flux) for the highly relativistic
configuration $\gamma v=2.0$ with spin $s=0.80$. The profile marks the
transient peak kick velocity $V_p$ immediately following the plunge and
merger, and the final stabilized velocity $V_f$, which share the same sign
with $|V_f|<|V_p|$: the ringdown radiation acts as a brake, so the antikick
$\Delta V$ is positive in this regime while the remnant's final
speed is reduced.
  }
\end{figure}

\begin{figure}[t]
  \includegraphics[width=\columnwidth]{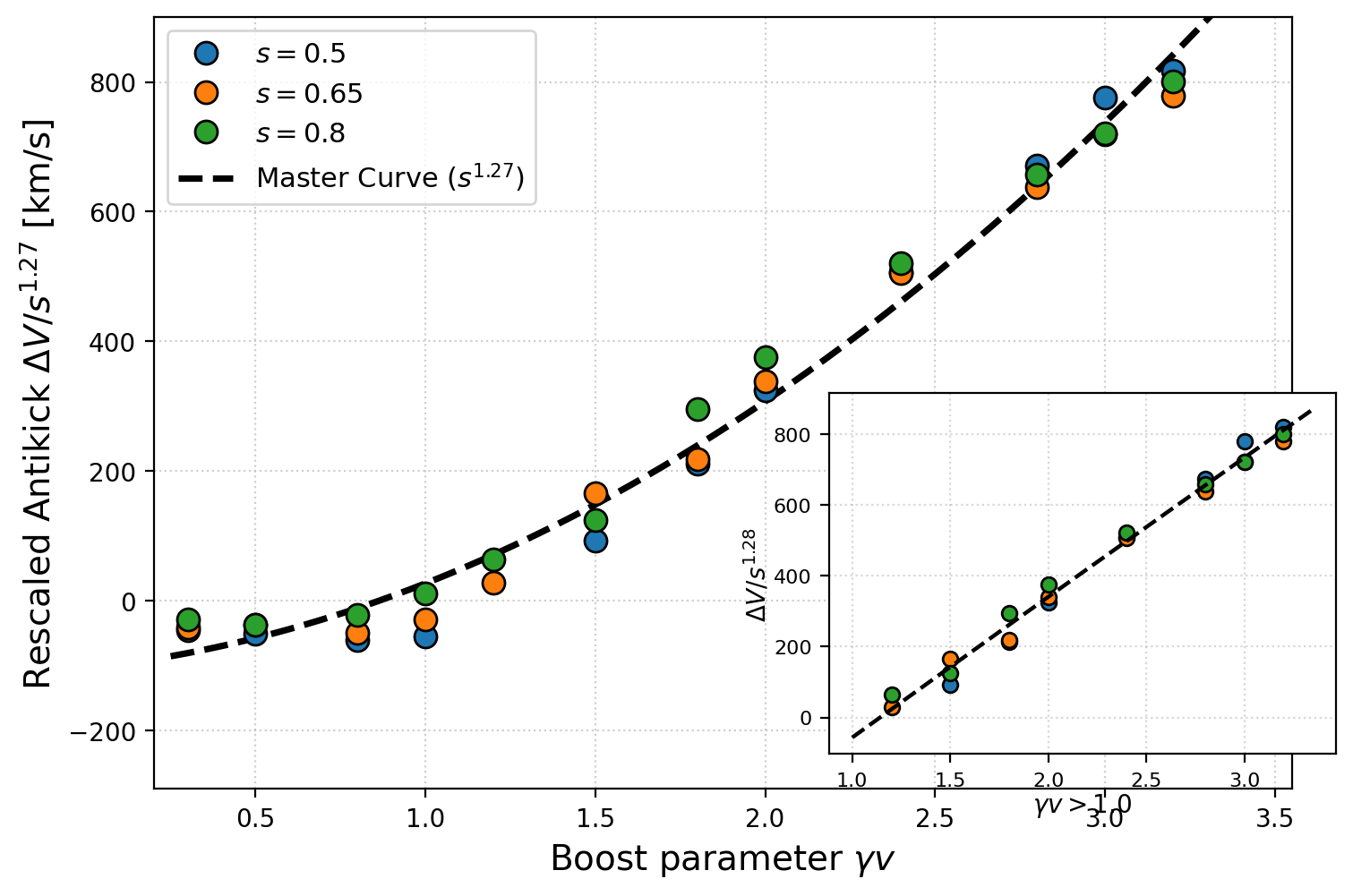}
  \caption{\label{fig:Scaling}
Main panel: The antikick velocity $\Delta V \equiv V_f - V_p$ (in km/s) rescaled by the empirically derived spin-dependence factor $s^{1.27}$, shown as a function of the initial boost parameter $\gamma v$. The rescaled numerical data for varying primary black hole spins ($s = 0.50$, $0.65$, and $0.80$) cleanly collapse onto a single universal master curve. A global non-linear least squares fit over the full velocity domain yields an optimal scaling exponent of $n=1.27\pm0.08$. The dashed black line indicates a global quadratic fit to the superposed data, demonstrating that this unified scaling relation effectively captures the spin-dependent recoil deceleration. Inset: The rescaled antikick velocities isolated strictly to the highly relativistic regime ($\gamma v > 1.0$). In this strong-field limit, the statistical fit for the exponent tightens to $n=1.28\pm0.06$, confirming the robustness of the $s^{1.27}$ scaling rule for high-energy head-on collisions.
}
\end{figure}

\begin{figure}[t]
  \includegraphics[width=\columnwidth]{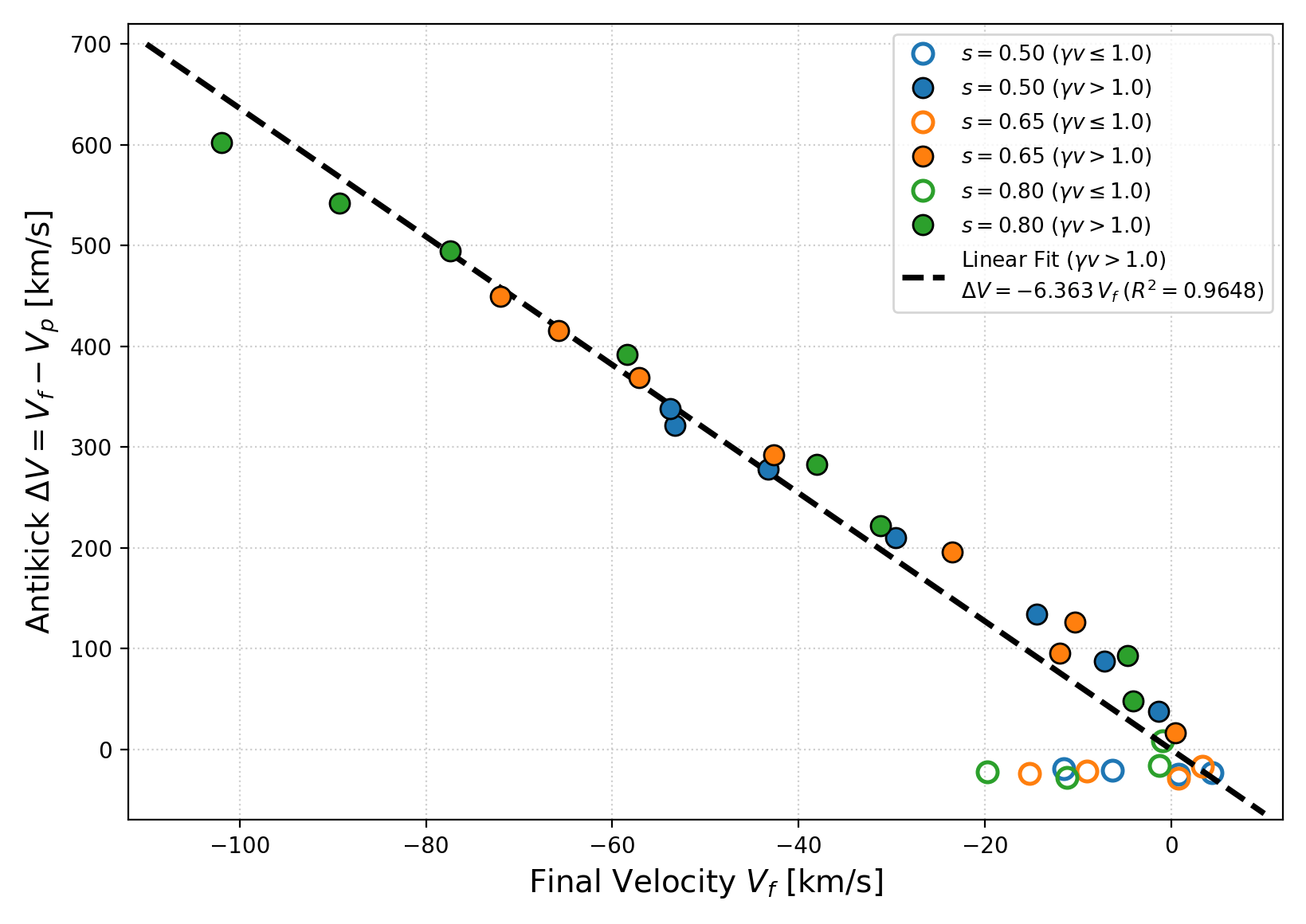}
  \caption{\label{fig:vp_vf}
The antikick velocity $\Delta V \equiv V_f - V_p$ (in km/s) plotted as a function of the final remnant recoil velocity $V_f$ for all simulated configurations. The dataset comprises a wide range of initial boost parameters $\gamma v \in [0.3, 3.2]$ across three primary dimensionless black hole spin magnitudes: $s = 0.50$ (blue circles), $s = 0.65$ (orange circles), and $s = 0.80$ (green circles). No spin-dependent rescaling or coordinate transformations are applied to the raw physical values. To distinguish between kinematic regimes, the mildly relativistic and transitional configurations ($\gamma v \le 1.0$) are shown as hollow circles, while the strictly relativistic sector ($\gamma v > 1.0$) is depicted by solid circles. The dashed black line represents a global linear regression ($\Delta V = m V_f$) applied exclusively to the $\gamma v > 1.0$ sector and rigidly constrained to pass through the origin ($\Delta V = 0$ at $V_f = 0$), enforcing the physical boundary condition that a vanishing final recoil necessitates a strictly zero antikick. In this strong-field domain, the data across all spin values tightly collapses onto a single proportional trajectory with a best-fit slope of $m=-6.363\pm0.137$ ($R^2=0.965$), i.e., $V_p=(1-m)V_f\approx7.4\,V_f$ (since $\Delta V = V_f - V_p$). This universal scaling relation confirms that for highly relativistic head-on mergers, the net momentum carried away by the asymmetrical ringdown radiation scales uniformly with the final asymptotic velocity of the remnant black hole, independent of the initial spin configuration.}
\end{figure}

The recoil velocities are collected in Tables~\ref{tab:kinematics_spin50}--\ref{tab:kinematics_spin80} and shown in
Fig.~\ref{fig:dv_gv}. For all spins the peak recoil $V_p$ rises with $\gamma v$,
changes sign near $\gamma v\simeq1.1$, and grows in magnitude at higher
momenta, reaching several hundred~km/s by $\gamma v=3.2$; the final recoil
$V_f$ tracks the same trend with much smaller amplitude, the difference being
the antikick. Figure~\ref{fig:vp_vf} shows that the antikick and the final recoil are
linearly correlated across all spins in the relativistic regime: a linear
fit through the origin to the $\gamma v>1$ data gives
$\Delta V=(-6.36\pm0.14)\,V_f$ with $R^2=0.96$, equivalently
$V_p\approx7.4\,V_f$ with the same sign, with only a weak residual
dependence on $s$. This near-universal proportionality is the central
empirical result of this work: once the common horizon forms, the ratio of
the final kick to the peak antikick is set by the post-merger relaxation
and largely ``forgets'' the initial momentum and spin, retaining only the
overall perturbation amplitude they produce. 
The spin dependence of the recoil \emph{amplitude} is less well determined.
A global fit of the form $\Delta V=s^{k}\,q(\gamma v)$, with $q$
quadratic, collapses the three spin sequences onto a single master curve
(Fig.~\ref{fig:Scaling}) with $k=1.27\pm0.08$ over the full range
($k=1.28\pm0.06$ restricting to $\gamma v>1$) and
$q(\gamma v)\simeq [75\,(\gamma v)^2+56\,(\gamma v)-104]$~km/s. This
empirical exponent should, however, be interpreted with care: per-momentum
power-law fits to only three spin magnitudes drift with $\gamma v$ (from
roughly linear toward quadratic at the highest momenta, and ill-defined
near $\gamma v\simeq1.1$ where the recoil changes sign), and all values lie
well below the close-limit prediction $s^3$ of Appendix~\ref{app:cl},
whose leading-order validity does not extend to these relativistic boosts.
Resolving this would require additional spin values, e.g., $s=0.683$ and
$0.890$ for equal spacing in $s^3$, and lower as well as higher initial
momenta $\gamma v$.

\subsection{Radiated energy and remnant mass}
The radiated energy $E_{\rm rad}/M_{\rm ADM}$ in
Tables~\ref{tab:kinematics_spin50}--\ref{tab:kinematics_spin80} grows
monotonically with $\gamma v$ and depends only weakly on the spin magnitude,
 mirroring the weak
spin dependence found for the grazing energy maximum~\cite{Healy:2024lhl}.
For the present head-on, momentum-limited sequence the radiated energy
reaches $\approx7\%$ by $\gamma v=3.2$, well below the grazing maximum of
$\approx32\%$~\cite{Healy:2024lhl} and the head-on nonspinning value of
$\approx13\%$~\cite{Healy:2015mla}, since the recoil-optimized configurations
studied here are not those that maximize the radiated energy.

\section{Discussion}
The most robust feature of the head-on recoil studied here is the
near-universal proportionality between the peak and the final kick,
$V_p\approx7.4\,V_f$, i.e., $\Delta V\approx-6.4\,V_f$
(Fig.~\ref{fig:vp_vf}), which holds across the full range of relativistic
momenta and spin magnitudes. Physically, this reflects that the recoil is
generated almost entirely after the common apparent horizon forms: the
distorted remnant rings down through its QNMs, and the ratio of
the transient peak to the residual final velocity is governed by that
relaxation rather than by the details of the progenitors. The QNM
model of Appendix~\ref{app:qnm}, built on the fundamental $\ell=2$ and $3$
Schwarzschild frequencies, reproduces qualitatively this peak-to-final structure, and the
close-limit/double-Kerr estimate of Appendix~\ref{app:cl} provides an
independent analytic handle on the same dynamics.

The spin dependence of the recoil amplitude is more subtle. The close-limit
computation isolates a surviving $\Re[A^{2,\pm2}\bar A^{3,\pm2}]$ coupling that
scales as $s\times s^2=s^3$, whereas the three spin magnitudes available here
do not yet pin down the empirical exponent, which appears closer to linear at
the highest momenta. While this close limit computation is set in a lower than relativistic
speeds regime, a denser spin sequence could also help to resolve this
exponent and to test whether the analytic $s^3$ prediction sets in only
asymptotically and what fraction contributes to the high energy collisions.

It is worth situating these head-on results against the grazing
configurations studied previously. There, the recoil reaches
$\sim28\,562$~km/s~\cite{Healy:2022jbh} and the radiated energy
$\sim32\%$~\cite{Healy:2024lhl}, both far larger than in the head-on case,
because orbital angular momentum and the associated zoom-whirl dynamics near
the critical impact parameter greatly enhance the asymmetric emission. The
head-on case isolates the spin-driven, post-merger contribution to the recoil
and is therefore the cleaner setting in which to study the relaxation
dynamics. Finally, as a motivation rather than a quantitative claim, the
high-energy collision regime is of broad interest through its connection to
the gauge/gravity duality, where bulk collisions map to the thermalization of
strongly coupled matter~\cite{Cardoso:2012qm,Berti:2010ce}.

\section{Conclusions}
We have characterized the recoil produced in high-energy head-on collisions
of equal-mass black holes with anti-aligned spins, for spin magnitudes
$s=0.5,\,0.65$, and $0.8$, and momenta $0.3\le\gamma v\le3.2$.
intrinsically spin-driven, post-merger effect: the velocity reaches its peak magnitude $V_p$ during the ringdown and relaxes to a final value $V_f$, with a near-universal relation $V_p\approx7.4\,V_f$
($\Delta V\approx-6.4\,V_f$) that is largely independent of momentum and
spin. We
have supported these numerical findings with three analytic descriptions---a
ZFL expansion, a QNM model of the antikick, and
a superposed boosted double-Kerr close-limit estimate---which together
predict a leading $s^3$ scaling of the recoil amplitude that remains to be
confirmed with additional spin values. These results complement the
previously established maxima of the grazing recoil and radiated
energy~\cite{Healy:2022jbh,Healy:2024lhl} and sharpen the picture of how
spin and relativistic momentum jointly determine the kick of the merger
remnant.

The ZFL computation of Appendix~\ref{app:zfl} provides the leading analytic
alternative to the close-limit description, valid in the opposite
(ultrarelativistic, instantaneous-collision) regime. In contrast to the
close-limit $s^3$ law, the ZFL momentum flux is linear in the spin,
$P_z\propto s$, with the momentum dependence of Eq.~(\ref{eq:Pzgv})
peaking at $\gamma v=1.90$, while the spin correction to the radiated
energy, Eq.~(\ref{eq:E2}), peaks at $\gamma v=4.25$. 
The empirical exponent $s^{1.27\pm0.08}$ found here lies between the ZFL linear and the close-limit
cubic predictions, suggesting that both channels contribute in the
relativistic regime; disentangling them requires the denser spin grid
discussed above. Extending the momentum range toward $v\to c$ with the
low-spurious-radiation initial data of Ref.~\cite{Ruchlin:2014zva}, and
comparing with recent studies of ultrarelativistic black-hole
encounters~\cite{Zhu:2026mhn}, will further reveal the spin dependence of
the recoil at the highest attainable energies.

\begin{acknowledgments}
COL gratefully acknowledges support from NSF awards AST-2319326,
PHY-2207920 and PHY-2513442.
Computational resources were also provided by the BlueSky
Clusters, Green Prairies, and White Lagoon at the Rochester Institute
of Technology, which were supported by NSF grants
No.\ PHY-1229173, No.\ PHY-1726215, and No.\ PHY-2018420.
This work also used the ACCESS computational resources from the
allocation PHY060027.
HN was supported by JSPS KAKENHI Grant Numbers JP21H01082, JP23K20845,
JP21K03582, JP23K03432 and JP26K07074, 
and would like to acknowledge the valuable support of 
the Research Centre for Relational Studies of Ryukoku University.
\end{acknowledgments}

\appendix
%

\section{Zero-frequency-limit analysis}\label{app:zfl}

The ZFL approximation was introduced by
Smarr~\cite{Smarr:1977fy} for the gravitational radiation emitted during the
high-energy scattering or collision of two black holes, and was shown to
provide an estimate of the total radiated energy, its spectrum, and its
angular distribution for high-velocity collisions of two particles in the
linearized approximation~\cite{Payne:1983rrr}. We extend the computation
of Ref.~\cite{Berti:2010ce} to include spin dependence in the head-on case.
Note that we rotate the coordinates so that the recoil is aligned with the $z$ axis in this Appendix.

The radiated energy and momentum are
\begin{align}\label{eq:Erad}
  \frac{d^2E_{\rm rad}}{d\omega\, d\Omega}
   &= 2\omega^2\!\left(T^{\mu\nu}T_{\mu\nu}^{*}
      -\tfrac12 T^{\lambda}{}_{\lambda}T^{*\kappa}{}_{\kappa}\right)\,,\\
  \label{eq:Prad}
  \frac{dP^{\rm rad}_i}{d\omega} &= \int d\Omega\,
     \frac{d^2E_{\rm rad}}{d\omega\,d\Omega}\, n_i \,,
\end{align}
with $n_i$ the unit vector in the $i$-direction. The stress tensor of a
spinning particle along the world line $\zeta^\alpha$~\cite{Mino:1995fm} is
\begin{equation}
  T^{\mu\nu}(x^\alpha)=\tfrac12\,\partial_\lambda\!\left[\int d\tau\,
   \big(u^\mu\Sigma^{\nu\lambda}+u^\nu\Sigma^{\mu\lambda}\big)\,
   \delta^4(x^\alpha-\zeta^\alpha)\right] \,,
\end{equation}
with the total momentum and spin tensor
$\Sigma^{\mu\nu}=- (1/M)\, \epsilon^{\mu\nu\rho\sigma}p_\rho S_\sigma$ and
$S_t=-S_i v^i$~\cite{Racine:2008kj}. 
Carrying out the $\tau\!\to\!t$ reduction, integrating by
parts, and dropping surface terms, the constant-spin encounter of two holes
yields a Fourier-space stress tensor that decomposes in Bessel functions
$J_n$ of the first kind:
\begin{align}\label{eq:Tfull}
\begin{split}
 T^{\mu\nu}(\mathbf{k},\omega)=&
 \sum_{k=1}^{2} \Biggl[ \frac{-k_\lambda u_k^\mu\Sigma_k^{\nu\lambda}}
   {4\pi\omega\gamma_k\,(1-\lambda_k v_k\sin\theta\cos\phi)}\\
 &\quad\times e^{-i\lambda_k\xi_k\omega\sin\theta\sin\phi}\\
 &-\frac{ik_\lambda\Sigma_k^{\nu\lambda}}{4\pi\gamma_k}
   \sum_{n=-\infty}^{\infty}e^{in\phi}
   J_n(-\lambda_k\sin\theta\,\xi_k\omega)\\ &\quad\times\int_0^{\infty}\!dt\,e^{i(\omega+n\Omega)t}u_k'^{\,\mu}(t) \Biggr]
   \;+\;\mu\leftrightarrow\nu \,,
\end{split}
\end{align}
where $u_k'^{\,\mu}\equiv du_k^{\mu}/dt$, so that in the ZFL the last
term reduces to the net velocity change of each hole, and we adopt the
$1/(2\pi)$ Fourier-transform normalization throughout (reflected in the
$1/(4\pi\gamma_k)$ prefactors).

\subsection{Head-on case}\label{app:zfl-headon}
For a head-on collision, the second (Bessel) term in Eq.~\eqref{eq:Tfull} drops out. 
Ending the integration at $t^{*}\!\approx\!2M/v$ when
the individual horizons touch (in the regime $v\sim c$), the angular energy
distribution reduces (in the $t^{*}\to0$ limit) to
\begin{equation}\label{eq:d2E}
\begin{split}
 \frac{d^2E}{d\omega\,d\Omega}=&\frac{v^2(x^2-1)}{4\pi^2(v^2x^2-1)^2}
   \Bigl\{ M^2v^2(x^2-1)\\
   &+2M^4s^2\omega^2\sqrt{1-v^2}\,(x^2-1)(v^4x^2-1)\cos2\phi\\
   &-2M^4s^2\omega^2\sqrt{1-v^2}\\
   &\quad\times\big[x^2(v^4(x^2+1)+4v^2+1)+1\big]\Bigr\} \,,
\end{split}
\end{equation}
where $x=\cos\theta$ is the cosine of the polar angle and $\phi$ the corresponding azimuthal angle from the collision
axis.
Integrating over the solid angle gives the total radiated energy,
\begin{align}\label{eq:B11}
\begin{split}
 E=&\frac{\omega_c}{36\pi v^3}\Big\{
   -3\ln\!\Big(\tfrac{2}{v+1}-1\Big)\big[3M^2v^2(v^4+2v^2-3)\\
   &-2M^4s^2\omega_c^2\sqrt{1-v^2}\,(v^6+3v^4-17v^2-3)\big]\\
   &-2\big[9M^2(v^2-3)v^3\\
   &\quad+2M^4s^2\omega_c^2\sqrt{1-v^2}(5v^4+54v^2+9)v\big]\Big\} \,,
\end{split}
\end{align}
where $\omega_c$ is an effective cutoff frequency,
and the radiated linear momentum expanded for small $t^{*}$,
\begin{equation}\label{eq:B12}
\begin{split}
 P_z=&\frac{2 M^3 s\, \omega_c^3 t^{*}\sqrt{1-v^2}}{45\pi v^2}\\
 &\times\big[-8v^5+75v^3+45(v^2-1)^2\tanh^{-1}v-45v\big] \,.
\end{split}
\end{equation}
Since $t^{*}\propto 1/v$, this is invalid near $v=0$; to leading order
\begin{equation}\label{eq:Pzv}
 P_z=\frac{32M^3 s\, \omega_c^3 v^2 x}{45\pi} \,,
\end{equation}
where $x$ is half the separation at which integration stops. 
Thus, $P_z$ is nonzero (unlike the nonspinning equal-mass case), 
$\propto S=s M^2$, and $\propto v^2$. 
For aligned spins and for spins along $x$, the net radiated
momentum vanishes in all directions.

\subsection{Explicit fitting formulas}\label{app:zfl-fits}
Substituting $v\to\gamma v/\sqrt{1+(\gamma v)^2}$ in Eq.~\eqref{eq:B12},
\begin{align}\label{eq:Pzgv}
 P_z=&\frac{2 M^3 s\, \omega_c^3 t^{*}}{45\pi}\,
 \frac{1}{(\gamma v)^2[1+(\gamma v)^2]^2} \cr
 & \times \Bigl\{
 22(\gamma v)^5-15(\gamma v)^3-45(\gamma v)
 \cr &
   +45\sqrt{1+(\gamma v)^2}\,\ln[\gamma v+\sqrt{1+(\gamma v)^2}] \Bigr\}
 \,,
\end{align}
which peaks at $\gamma v=1.896396$ ($v=0.884553$). Splitting
$E_{\rm rad}=E_0+E_2 s^2$,
\begin{equation}\label{eq:E0}
\begin{split}
 E_0=\frac{M^2\omega_c}{2\pi}\!\Bigg\{&\frac{3+2(\gamma v)^2}{1+(\gamma v)^2}\\
   &-\frac{[3+4(\gamma v)^2]\ln[\gamma v+\sqrt{1+(\gamma v)^2}]}
     {(\gamma v)[1+(\gamma v)^2]^{3/2}}\Bigg\} \,,
\end{split}
\end{equation}
which approaches $M^2\omega_c/\pi$ as $\gamma v\to\infty$, and
\begin{equation}\label{eq:E2}
\begin{split}
 E_2=&\frac{M^4\omega_c^3}{36\pi}\Bigg\{
   -\frac{4[9+72(\gamma v)^2+68(\gamma v)^4]}{(\gamma v)^2[1+(\gamma v)^2]^{3/2}}\\
   &+\frac{12[3+2(\gamma v)^2][1+8(\gamma v)^2+8(\gamma v)^4]}
     {(\gamma v)^3[1+(\gamma v)^2]^2}\\
   &\quad\times\ln[\gamma v+\sqrt{1+(\gamma v)^2}]\Bigg\} \,,
\end{split}
\end{equation}
which peaks at $\gamma v=4.246248$ ($v=0.973372$). 

\begin{figure}[t]
  \includegraphics[width=\columnwidth]{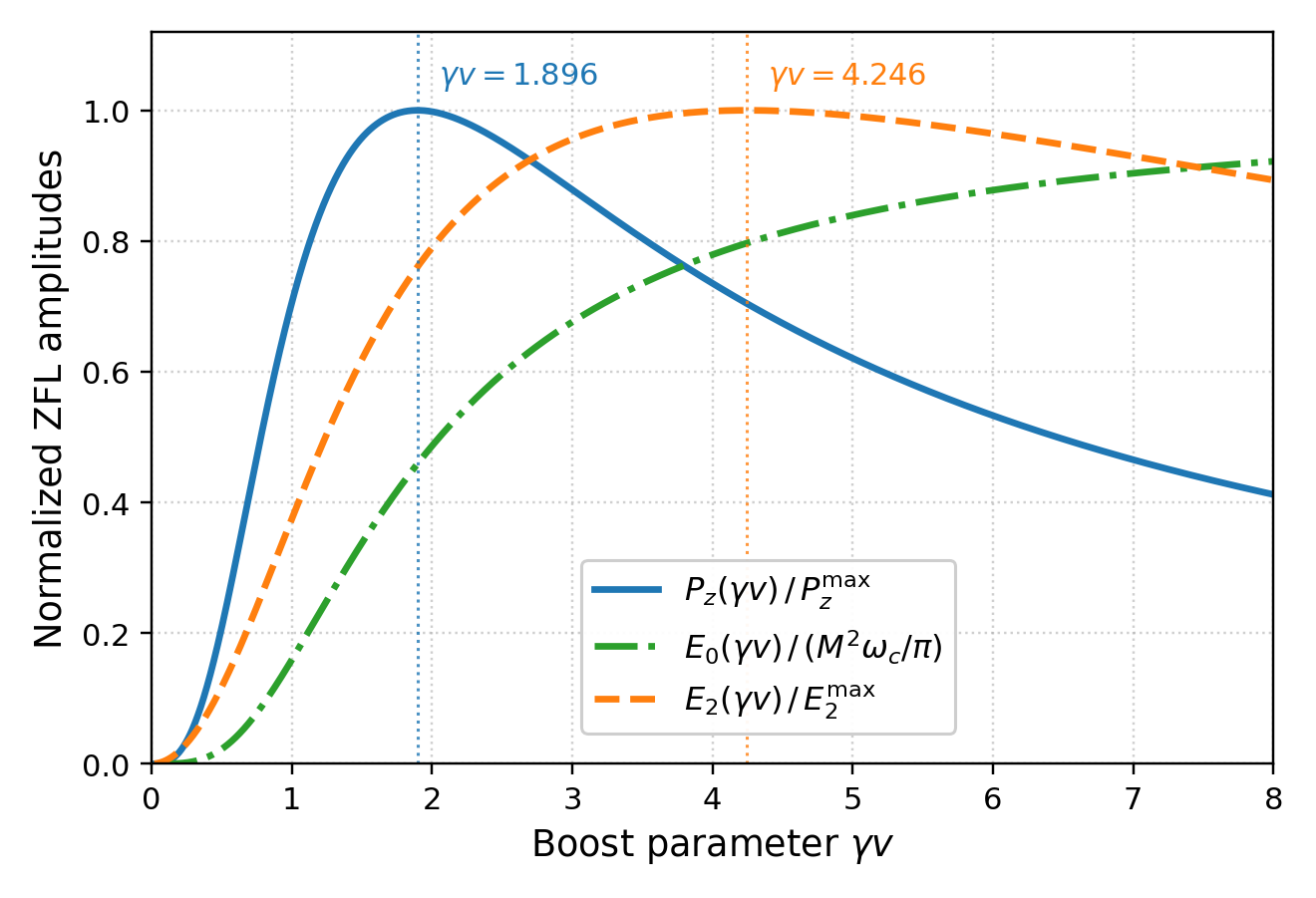}
  \caption{\label{fig:zfl_shapes}
Normalized $\gamma v$ dependence of the ZFL amplitudes. The radiated
linear momentum $P_z$ of Eq.~(\ref{eq:Pzgv}) (solid blue, at fixed
$t^{*}$ and $\omega_c$, normalized to its maximum) peaks at
$\gamma v=1.896$ ($v=0.885$); the spin-quadratic correction to the
radiated energy $E_2$ of Eq.~(\ref{eq:E2}) (dashed orange, normalized to
its maximum) peaks at $\gamma v=4.246$ ($v=0.973$); and the nonspinning
radiated energy $E_0$ of Eq.~(\ref{eq:E0}) (dash-dotted green) grows
monotonically toward its asymptote $M^2\omega_c/\pi$. Dotted vertical
lines mark the two maxima.}
\end{figure}

The three $\gamma v$-dependent shape functions are compared in
Fig.~\ref{fig:zfl_shapes}: the radiated momentum $P_z$
[Eq.~(\ref{eq:Pzgv}), at fixed $t^{*}$ and $\omega_c$] and the spin
correction to the radiated energy $E_2$ [Eq.~(\ref{eq:E2})] are
normalized to their maxima, while the monotonic nonspinning energy $E_0$
[Eq.~(\ref{eq:E0})] is normalized to its asymptotic value
$M^2\omega_c/\pi$. The recoil-generating channel peaks at markedly lower
boost ($\gamma v=1.90$) than the spin correction to the energy
($\gamma v=4.25$), bracketing the momentum range
$0.3\le\gamma v\le3.2$ covered by the simulations of the main text.

\section{Perturbative model of the spin-dependent recoil}
\label{app:recoil}

This Appendix collects the analytic results underlying the recoil model,
\begin{equation}
  A(\gamma v,s)\;\simeq\; C\,(\gamma v)^{n}\,s^{k} \,,
  \label{eq:Ansatz}
\end{equation}
used in the main text. Here, $A$ is the amplitude of the post-merger
linear-momentum flux $dP^{z}/dt$, ad $\gamma v$ and  $s=a/M$ denote 
the boost and the (dimensionless) hole spin, respectively. 
We first model the antikick as a beat of the
two leading QNMs of the remnant
(Appendix~\ref{app:qnm}), and then derive the cubic spin scaling $k=3$ from
the close-limit data of two boosted, oppositely spinning Kerr holes
(Appendix~\ref{app:cl}). Throughout, we set $G=c=1$ and use units $M=1$ for
the remnant mass where convenient.

\subsection{Antikick as a quasinormal beat}
\label{app:qnm}

Essentially all of the recoil is generated after the holes merge, when the
remnant rings down as a perturbed Schwarzschild black hole
\cite{Schnittman:2007ij}. The linear-momentum flux couples radiative
multipoles of adjacent $\ell$, so its leading post-merger behavior is set
by the interference (beat) of the fundamental $\ell=2$ and $\ell=3$
gravitational ($s=2$, overtone $n=0$) QNMs. Writing each mode as
$e^{-i\omega_{\ell}(t-t_{0})}$ with $\omega_{\ell}=\omega_{R}^{(\ell)}
-i\,\omega_{I}^{(\ell)}$ and the ringdown onset $t_{0}$, the surviving
cross term is the damped beat
\begin{equation}
\begin{split}
  \frac{dP^{z}}{dt}
  &=\mathrm{Re}\!\left[A\,e^{-i(\Delta-i\Gamma)(t-t_{0})}\right]\\
  &=|A|\,e^{-\Gamma(t-t_{0})}\cos\!\big[\Delta(t-t_{0})-\arg A\big] \,,
\end{split}
  \label{eq:beat}
\end{equation}
with the beat frequency and combined decay rate,
\begin{equation}
  \Delta\equiv\omega_{R}^{(3)}-\omega_{R}^{(2)} \,,
  \qquad
  \Gamma\equiv\omega_{I}^{(2)}+\omega_{I}^{(3)} \,.
  \label{eq:DeltaGamma}
\end{equation}
Using the fundamental Schwarzschild values
$M\omega^{(2)}=0.37367-0.08896\,i$ and
$M\omega^{(3)}=0.59944-0.09270\,i$, we have
\begin{equation}
  M\Delta=0.2258 \,, \quad M\Gamma=0.1817 \,, \quad \Gamma/\Delta=0.81 \,,
\end{equation}
i.e., the beat is strongly damped ($\Gamma\!\sim\!\Delta$): the integrated
momentum $P^{z}(t)=\int_{t_{0}}^{t}(dP^{z}/dt')\,dt'$ exhibits a single
overshoot---the antikick---before relaxing to the final recoil $P^{z}_{\rm
fin}$. The overshoot
\begin{equation}
  \Delta P\equiv P^{z}_{\rm peak}-P^{z}_{\rm fin} \,,
\end{equation}
is invariant under the trigger time $t_{0}$, which enters
Eq.~\eqref{eq:beat} as a pure time translation; it therefore fixes $|A|$
independently of the (gauge-dependent) ringdown onset. For in-phase
excitation ($\arg A=0$), one finds the configuration-independent numbers,
\begin{equation}
  P^{z}_{\rm fin}=2.16\,|A| \,, \qquad \Delta P=0.76\,|A| \,,
\end{equation}
i.e., an antikick fraction $\Delta P/P^{z}_{\rm fin}\simeq0.35$. 
These in-phase numbers illustrate the mechanism; the much larger ratio
$\Delta P/P^{z}_{\rm fin}\approx6$--$7$ observed in the numerical
simulations (Fig.~\ref{fig:vp_vf}, where $P_{\rm peak}\approx7.4\,
P_{\rm fin}$) corresponds to excitation substantially out of phase,
$\arg A\neq0$, for which the final momentum results from a
near-cancellation while the overshoot does not. Reading
$\Delta P$ off a numerical waveform thus determines the amplitude
$A=A(\gamma v,s)$ of Eq.~\eqref{eq:Ansatz} without fitting $t_{0}$.

\subsection{Close-limit origin of the cubic spin dependence}
\label{app:cl}

To fix the exponents in Eq.~\eqref{eq:Ansatz} we treat the late-time
configuration as a single distorted Schwarzschild hole, obtained by
superposing two equal-mass Kerr holes with equal and opposite spins and a
relative boost, in the close limit. This generalizes the static,
equal-and-opposite-spin construction of Ref.~\cite{Khanna:2001ch} (its
Case~II) by including the boost; the superposition is performed at the
level of the full Kerr data rather than the conformally flat Bowen--York
form, and the resulting perturbation is decomposed into the even (polar)
and odd (axial) Regge--Wheeler--Zerilli master functions
$\psi^{\rm e}_{\ell, m}$ and $\psi^{\rm o}_{\ell, m}$.

\paragraph*{Momentum flux and selection rules.}
With the radiative multipoles $A_{\ell, m}$ of the (time-integrated) Weyl
scalar expanded in spin-weight ($-2$) harmonics, the axial recoil reads
\cite{Ruiz:2007yx} (see also Sec.~8.9.3 of
Ref.~\cite{AlcubierreBook2008})
\begin{equation}
\begin{split}
  \frac{dP^{z}}{dt}=\lim_{r\to\infty}\frac{r^{2}}{16\pi}
  \sum_{\ell, m}\bar A_{\ell, m}\big(&a_{\ell, m}A_{\ell, m}
  +b_{\ell, m}A_{\ell-1,m}\\
  &+b^{*}_{\ell+1,m}A_{\ell+1,m}\big) \,,
\end{split}
  \label{eq:dPz}
\end{equation}
where
\begin{align}
  a_{\ell, m}&=\frac{m}{\ell(\ell+1)} \,, \\
  b_{\ell, m}&=\frac{1}{2\ell}
  \sqrt{\frac{(\ell-2)(\ell+2)(\ell-m)(\ell+m)}{(2\ell-1)(2\ell+1)}} \,.
\end{align}
The equal-and-opposite-spin configuration is symmetric under reflection
through the orbital plane, which enforces
$\,|A_{\ell,m}|=|A_{\ell,-m}|$. Because $a_{\ell, m}\propto m$ is odd in
$m$, every diagonal term cancels in the $\pm m$ sum,
$a_{\ell, m}\!\left(|A_{\ell, m}|^{2}-|A_{\ell,-m}|^{2}\right)=0$, and the
only surviving contribution is the adjacent-$\ell$ cross term
$\propto\mathrm{Re} \big[\bar A_{3,\pm2}A_{2,\pm2}\big]$. Writing
$A_{\ell, m}=\psi^{\rm e}_{\ell, m}+i\,\psi^{\rm o}_{\ell, m}$ and summing
over $m=\pm2$,
\begin{equation}
  \frac{dP^{z}}{dt}\;\propto\;
  \underbrace{\psi^{\rm e}_{3,2}\,\psi^{\rm e}_{2,2}}_{\text{mass--mass}}
  \;+\;
  \underbrace{\psi^{\rm o}_{3,2}\,\psi^{\rm o}_{2,2}}_{\text{current--current}} \,,
  \label{eq:channels}
\end{equation}
the two parities contributing through the mass--mass and current--current
channels, respectively. (The mixed term $\psi^{\rm e}\psi^{\rm o}$ is
imaginary and drops out of the real part; for the pure static
configuration of Ref.~\cite{Khanna:2001ch} only $\ell=2$ is excited and
both channels vanish, so there is no recoil at all.)

\paragraph*{Spin content of the multipoles.}
Projecting the boosted, oppositely spinning close-limit data onto the
master functions gives
\begin{align}
  \psi^{\rm e}_{2,\pm2}&=\mathcal{O}(a^{0}), &
  \psi^{\rm e}_{3,\pm2}&\equiv0, \nonumber\\
  \psi^{\rm o}_{2,\pm2}&=\mathcal{O}(a^{1}), &
  \psi^{\rm o}_{3,\pm2}&=\mathcal{O}(a^{2}) \,,
  \label{eq:spinpowers}
\end{align}
to leading order in the spin.
The spin parity of each multipole under spin reversal follows the rule 
$\psi^e_{\ell,m}(-a) = (-1)^\ell \psi^e_{\ell,m}(a)$ and 
$\psi^o_{\ell,m}(-a) = (-1)^{\ell+1} \psi^o_{\ell,m}(a)$, i.e., a manifestation 
of the $a \to -a$ asymmetry of the configuration: polar multipoles are even 
in $a$ and axial multipoles odd in $a$ at $\ell = 2$, with the alternation 
reversing at $\ell = 3$.
Two features of Eq.~\eqref{eq:spinpowers} are
decisive. First, the polar octupole vanishes identically,
$\psi^{\rm e}_{3,\pm2}\equiv0$ (consistent with the fact that no polar
$\ell=3$ is generated without a parity-breaking distortion, cf.\ the
odd-parity ``twist'' data of Ref.~\cite{Baker:1999sj}); the mass--mass
channel of Eq.~\eqref{eq:channels} is therefore absent. Second, the axial
octupole $\psi^{\rm o}_{3,\pm2}$ turns on only at $\mathcal{O}(a^{2})$ and
only for $m=\pm2$ (the axial $\ell=3$, $m=0$ projection is zero), so it
exists solely in the channel that feeds the recoil.

\paragraph*{Cubic spin scaling.}
Combining Eqs.~\eqref{eq:channels} and \eqref{eq:spinpowers}, the radiated
recoil arises entirely through the current--current channel,
\begin{equation}
  \frac{dP^{z}}{dt}\;\propto\;
  \psi^{\rm o}_{3,\pm2}\,\psi^{\rm o}_{2,\pm2}
  \;=\;\mathcal{O}(a^{2})\times\mathcal{O}(a^{1})
  \;=\;\mathcal{O}(a^{3}),
\end{equation}
i.e., the current octupole ($\ell=3$ axial, and $\propto a^{2}$) beating
against the current quadrupole ($\ell=2$ axial, and $\propto a^{1}$). Hence
$k=3$. The result is odd in $a$, as required by the $a\!\to\!-a$
symmetry, and the would-be spin--orbit term $k=1$ is excluded: the
assembled near-zone expression does contain an $\mathcal{O}(a)$
contribution, but it falls as $1/r^{2}$ and carries no radiative
($1/r$) part, so it transports no momentum to null infinity. The leading
\emph{radiated} recoil is therefore genuinely cubic in the spin.

\paragraph*{Explicit close-limit flux.}
Evaluating Eq.~\eqref{eq:dPz} with the Regge--Wheeler--Zerilli mode
functions of the superposed boosted double-Kerr data gives, to
$\mathcal{O}(v^{2})$,
\begin{equation}\label{eq:dPzDK}
\begin{split}
 \frac{dP_z}{dt}=&-\frac{a^3M\,\Delta(t)\,(15M+2r)\,V(t)\,\dot V(t)}{6300\,r^5}\\
 &-\frac{a^3M\,\Delta(t)\,(15M+2r)\,\dot V(t)\,\ddot V(t)}{25200\,r^3} \,,
\end{split}
\end{equation}
where $\Delta(t)$ and $V(t)=\dot\Delta(t)$ are the separation and relative
velocity, respectively, making the overall $a^{3}$ prefactor and the velocity-quadratic
($V\dot V$) structure fully explicit. Modeling $\Delta(t)$ by a radial
Schwarzschild geodesic near the horizon,
\begin{equation}
V= -\dfrac{\Delta}{2M} \,, \quad
\dot V= \dfrac{\Delta}{4M^{2}} \,, \quad 
\ddot V= -\dfrac{\Delta}{8M^{3}} \,,
\end{equation}
yields the
order-of-magnitude estimate
\begin{equation}\label{eq:dPzDK2}
 \frac{dP_z}{dt}=\frac{a^3\,\Delta(t)^3\,(15M+2r)(16M^2+r^2)}{806400\,M^4 r^5} \,,
\end{equation}
evaluated at finite observer radius $r$; the radiative content must
ultimately be read off at null infinity, so Eq.~\eqref{eq:dPzDK2} fixes
the scalings rather than the coefficient $C$.

\paragraph*{Velocity scaling.}
The entire perturbation is sourced by the boost: each radiative multipole
is built from the velocity profile $V(t)$ and its time derivatives and
vanishes when the boost is switched off. The recoil is thus quadratic in
the radiative amplitude and, to leading order in the low-velocity
expansion---in agreement with the post-Newtonian momentum flux, which
requires a velocity factor to be nonzero \cite{Racine:2008kj}---scales as
$(\gamma v)^{2}$, so $n=2$.

\paragraph*{Summary.}
Equations~\eqref{eq:channels} and \eqref{eq:spinpowers} fix the exponents of
the recoil model \eqref{eq:Ansatz} to
\begin{equation}
  {\,A(\gamma v,s)\;\simeq\;C\,(\gamma v)^{2}\,s^{3}\,};
  \qquad n=2,\quad k=3,
\end{equation}
with $k=3$ derived here from the multipolar content of the close-limit
data (current quadrupole $\times$ current octupole) rather than fit, and
the linear-in-spin contribution shown to be non-radiative. An equivalent
configuration rotated so that the recoil lies along the symmetry ($z$)
axis is best suited to this mode analysis (cf. the rotations in the
appendix of Ref.~\cite{Campanelli:2008nk}). Because the cubic law is
established analytically, numerical-relativity runs spaced uniformly in
$s^{3}$ (e.g., $s=0.5,\,0.683,\,0.8$, and $0.890$) sample the fit on an even
grid and refine the coefficient $C$ rather than disentangle competing
powers.

\bibliographystyle{apsrev4-2}
\bibliography{references}

\end{document}